# OPTICAL MEMS DESIGN FOR TELECOMMUNICATIONS APPLICATIONS

**Vladimir A. Aksyuk, Maria E. Simon, Flavio Pardo, Susanne Arney**
Lucent Technologies, Bell Labs
Murray Hill, NJ  07479

**Daniel Lopez**
Agere Systems
Murray Hill, NJ  07479

**Anita Villanueva**
Massachusetts Institute of Technology
Cambridge, MA 02139

## ABSTRACT

As optical telecommunication networks become more complex, there is an emerging need for systems capable of very complex switching and manipulation of large numbers of optical signals. MEMS enable these systems by combining excellent capabilities and optical properties of macroscopic optomechanics with dense integration of multiple actuators on a single chip. Such optical MEMS present common design and process challenges, such as multiple electrical and optical IO, optical surface quality, optical integration density (fill factor) and actuator performance and reliability. We have used general design approaches such as pure-flexure design, electrostatic actuation and residual stress engineering in addressing these challenges. On several examples in this paper we illustrate these approaches along with underlying design tradeoffs and process requirements. We also describe specific numerical techniques useful for electrostatic actuator optimization and for analyzing the effects of residual stress.

## INTRODUCTION

Even during the current slowdown in the telecommunication industry the drive towards growth and sophistication continues in both core and metropolitan optical networks. With hundreds of wavelength channels being transmitted through a single fiber at ever-higher bit rates, flexibility and the ability to reconfigure, provision and restore connections through a network are critical. Many of the light manipulation tasks associated with such networks are cheaper and easier if performed optically. These applications are not limited to optical switching, but also include tasks such as dynamic gain equalization and dispersion compensation. These tasks have to be performed with very little optical penalty.

The individual operations can be performed very well by conventional optomechanical components. However, they are bulky, expensive and slow. More importantly, the number of signals that need to be manipulated is very large, often hundreds or even thousands. MEMS allow integration densities sufficient to deal with this large number of signals. Most optomechanical components can be scaled down using MEMS technologies without an increase in optical loss or other optical penalties. This technology enables new lightwave systems for larger, more flexible and more complex optical networks of the future.

The designer is typically challenged to achieve the densest integration in the optical domain (highest "fill factor" in a 1D or 2D array of MEMS devices). This requirement is universal for applications ranging from large optical switches to adaptive optics, from tunable gratings to WDM gain equalizers and Add-Drop multiplexors. While increasing the density of optical elements one has to continue to provide ways of individually actuating the devices.

The individual devices may be provisionally divided into an actuator and an optical element. Many optical elements have been demonstrated previously, including mirrors, filters, refractive and diffractive lenses, and polarization control elements [1-3]. In this paper instead of considering various optical elements, which are often specialized for a particular application, we will concentrate on actuators, specifically the aspects of actuator design common to a variety of optical MEMS.

For the dense integration and the large number of degrees of freedom that have to be independently controlled on a single chip, electrostatic actuation presents significant advantages over other actuation principles. It provides very low power dissipation, allowing more devices to be put closer together. The electric fields can be highly localized by design, leading to very low cross-talk between neighboring devices. Requiring no special materials, it is suitable for a variety of fabrication processes. The higher actuation voltage typically associated with electrostatic actuation is usually a reasonable price to pay for the benefits. Besides, unconventional electrostatic design approaches can provide ways to further reduce the driving voltages, as discussed below.

Although beyond the scope of this paper, integrated sensors and feedback can further improve electrostatic actuators, increasing range and decreasing voltage at the expense of the additional complexity of the sensor and the feedback electronics. This is in most cases feasible only if electronic circuits are closely integrated with the MEMS devices.

In this paper we will use several examples of optical MEMS devices to illustrate the common recurrent themes in device design. These are elastic elements and pure-flexure design (with no mechanical surface contact between moving parts), electrostatic actuation principle and residual stress engineering. We will attempt to show different ways these themes can manifest themselves and different approaches that can be used in understanding, modeling and utilizing them.

Very often a MEMS designer is confronted by nonlinear phenomena. Stress, residual or induced, leads to nonlinear mechanics, such as buckling. In fact, for both deforming beams and plates, the linear, small-displacement theory is valid only as long as the tension or compression along the beam or within the plate is small [4]. In the vast majority of cases capacitances in electrostatic actuators are nonlinear functions of displacements, producing position-dependent actuation forces and thus nonlinear equations of motion.

In some cases these nonlinearities are to be avoided, e.g., by strain-relieving elastic suspensions. In others they have to be dealt with, such as in designing electrodes for a tilting mirror. They can also be put to productive use and even enable new types of mechanisms. In all cases numerical modeling becomes very useful and necessary in analyzing the effects.

In our first example below we consider a simplified model of a beam-steering micromirror with an elastic suspension and electrostatic actuation by a fixed electrode. We describe a combination of numerical and analytical techniques useful in predicting the actuator performance characteristics. We also illustrate the modeling of residual stress effects on the mirror suspension, and show a useful analytical approach for spring optimization. Although the model we use to illustrate the techniques is very simple, the same techniques are applicable to a whole class of much more complex devices.

In another example of a tilting mirror a small displacement is transduced into comparably large tilt of a reflector. Achieving the required angle with a pure-flexure design presents a certain challenge. This device provides another example for the spring optimization.

The next part of the paper illustrates how residual stress can be used to an advantage. First, we show a mechanical part with deliberately built-in residual stress employed to create self-assembling devices, and then show residual stress enabling a completely different type of a tilting mirror device. This interesting device illustrates where the previously described analysis techniques are no longer applicable. It also takes the next step beyond the pure-flexure micromachines. The concept behind this device allows decoupling of actuation range and voltage, showing a way to a significant actuation voltage reduction.

Finally we consider the effects of the residual stress on the shape of 2D plates. Here we consider bimorph plates with large stress mismatch between layers. Such structures present an interesting case of geometrical mechanical nonlinearity and can be used for building a new type of a bi-stable (latchable) actuator.

## SIMPLE MIRROR – NUMERICAL ANALYSIS

In this section a self-consistent approach to solve the mechanical and electrostatic behavior of certain types of MEMS devices is presented. The approach combines numerical simulation with analytical formulas in order to reduce the total numerical simulation time. It works well and provides significant computational advantage when electrostatic and mechanical problems can be effectively decoupled.

Suppose we can postulate *a priori* that electrostatic forces influence only a few mechanical degrees of freedom (DOF). This is the case when several parts connected by springs move as separate rigid bodies (without deformation) under the influence of electrostatic forces, and the electrostatic forces are not acting on the springs. Such electrostatic forces can be calculated for all values of mechanical DOF and then the simple mechanical problem can be solved in the presence of these applied position-dependent forces. Moreover, when changes in some DOF are small, they can be treated as a perturbation.

We will illustrate the technique on a simple example below. Later in this paper we will give another example of an electromechanical system which is strongly coupled and for which this technique cannot be applied.

Figure 1 shows a typical 1-axis tilting mirror design. The electrically grounded mirror tilts around the $\hat{y}$ axis as a voltage $V$ is applied to the electrode. We will assume all the mechanical deformation occurs in the springs, the mirror is assumed rigid. The mechanical energy is given by,

$$E_M = \sum_i \frac{1}{2} K_i (r_i)^2 + \frac{1}{2} \tau \theta^2, \qquad Eq.\ 1$$

where $r_i$ is the displacement of the center of mass of the mirror from equilibrium $(i=x,y,z)$, $\theta$ is the mirror tilting angle, $\tau$ and $K_i$ are the torsional and linear spring stiffnesses respectively.

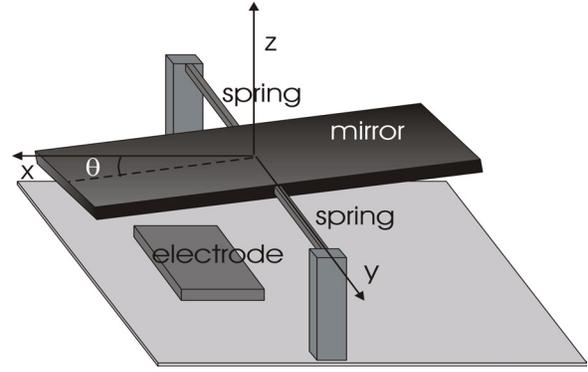

*Figure 1. Schematic of a 1-axis tilting mirror.*

The mirror is designed to have the torsional stiffness much smaller than the linear ones, in other words the lowest vibration eigenmode of the model is rotation. In this limit the linear terms (first term in Eq. 1) can be treated as a perturbative correction of the pure torsional model (last term in Eq.1).

The electrostatic energy is given by $E_E = \frac{1}{2} V^2 C$,

where $C$ is the capacitance and $V$ is the applied voltage on the electrode, and any other conductor is grounded. As the mirror was assumed rigid, the capacitance depends only on the mirror coordinates, $C(\theta, \vec{r})$. This capacitance can be obtained using any finite element electrostatic solver, such as the commercial CoventorWare™ software package. From the mechanical and electrostatic energies, the following set of equilibrium equations can be derived:

$$\tau \theta = \frac{1}{2} V^2 \frac{\partial C}{\partial \theta} \qquad Eq.\ 2$$

$$K_i r_i = \frac{1}{2} V^2 \frac{\partial C}{\partial r_i} \qquad Eq.\ 3$$

For the known function $C(\theta, \vec{r})$ these equations determine the equilibrium voltage $V(\theta)$ at which a given angle is attained.

The problem can be solved by a perturbative approach. The solving mechanism is:

**1. Solution of the unperturbed system.** Only the pure torsional model (last term in Eq.1) is considered. Numerical simulations are performed to get $C^0(\theta,0)$. Eq. 2 is solved to get the unperturbed solution $V^0(\theta)$.

**2. Correction of the trajectory $\vec{r}^0(\theta)$.** This correction is obtained solving Eq.3 using the unperturbed $V^0$ and $C^0$. To solve Eq.3, more numerical simulation is needed in order to evaluate the derivatives: $C^0(\theta, \delta_x)$ is needed to obtain

$$\frac{\partial C}{\partial x}(\theta,0) = \frac{C^0(\theta, \delta_x) - C^0(\theta,0)}{\delta_x}.$$

**3. Correction of the voltage $V^1(\theta)$.** Numerical simulations are performed to get $C^1(\theta, \vec{\rho}^0(\theta))$. Eq.2 is solved to get $V^1(\theta)$.

**4. Iteration of steps 2 and 3 until the trajectories converge.** $V^1$ and $C^1$ are used to recalculate $\vec{\rho}^1(\theta)$ (step 2). These new trajectories can be used to recalculate $V$ and $C$ (step 3) and so on. This procedure can be repeated until $\vec{\rho}^n \approx \vec{\rho}^{n+1}$.

Figure 2 shows voltage vs. tilt angle curves for a mirror, as calculated using this procedure. The voltage $V^0$ of a pure rotation and the voltage of a corrected model including displacements in $z$ are shown. The solution shown by the second curve took only a few iterations to converge. The modeled device was built, and experimental data are also shown. The corrected solution reproduces the experimental data very well. Voltages and angles are normalized to the snap-down values.

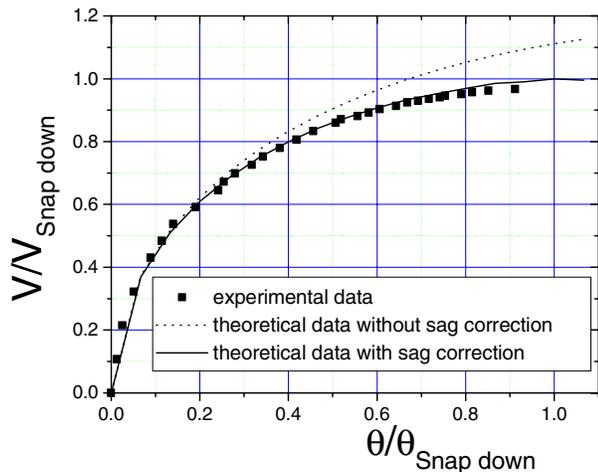

*Figure 2.* Voltages vs. angle curves. The dotted line corresponds to the unperturbed solution and the solid line corresponds to the corrected solution. The squares are the actual experimental data.

This method is most useful for more complex mechanical models with more degrees of freedom. For example it is very useful in analyzing gimbaled mirror models with 2 movable rigid bodies (mirror and gimbal ring) and 4 spring elements. The model contains 8 mechanical degrees of freedom – two tilt angles and 6 translations for the two bodies. For any given trajectory in the 2D angle space the two voltages can be calculated, taking into account the $x$, $y$ and $z$ displacements as perturbations. The same technique can also be easily adapted to analyze the effects of low-frequency mechanical vibrations on the device.

The technique provides significant computational savings for these types of problems since only a few capacitance calculations are required to obtain a trajectory. To compare, if standard iterative fully coupled electrostatic-mechanical method were used, multiple iterations of electrostatic and mechanical calculations would have been required for each point. Moreover, the number of iterations required usually diverges at the instability points, while our technique is completely insensitive to this issue.

## SIMPLE MIRROR – STRESS EFFECTS

The previous analysis assumed that the suspension springs are elastic with certain stiffness for rotation and displacement. However, the designer has to make sure those numbers are well-defined. Processing, packaging and thermal mismatch are only a few sources known to introduce stress in the mirror layer. A robust design should be insensitive to such stresses.

The design in Fig. 1 uses straight torsion members rigidly fixed on the sides. It is not strain-relieved. A compressive stress with such a suspension will lead to buckling. Buckling is a situation when the mechanical structure deforms, spontaneously breaking its symmetry even when no external forces are applied.

Almost any elastic-mechanical solver that can calculate the resonance modes and frequencies can easily predict the stress level, at which such deformation occurs. Fundamentally, a resonance frequency is a measure of energy required for a deformation of a particular shape. When spontaneous deformation occurs, such as in case of buckling, that means no energy is required for such deformation, and consequently the corresponding resonance frequency reaches 0 at this point.

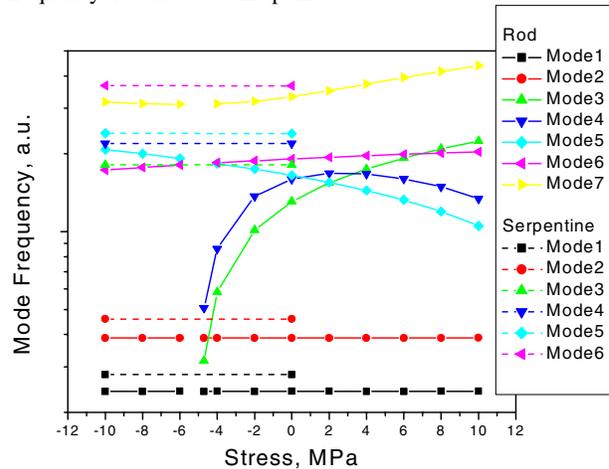

*Figure 3.* Computer-simulated resonance mode frequencies for double-gimbaled mirrors as a function of residual stress. "Serpentine" spring design is insensitive to stress, while "Rod" design exhibits buckling at –5 MPa. Compressive stress is negative.

Figure 3 shows a set of resonance frequencies of a 2D beam-steering micromirror device as a function of residual stress. Two sets of curves are shown. One set is for a device with properly designed, strain relieving "serpentine" torsional springs. This set shows resonance frequencies completely independent of the residual stress value. The other set of curves shows the same device when simple straight torsion "rods" with similar torsional spring constants were substituted for the original springs.

One can observe that frequencies for two mechanical resonance modes (both corresponding to in-plane translations) turn to 0 at about –5MPa. This means that above this stress level the mirror would buckle, i.e., spontaneously shift in the plane from its symmetrical configuration. This quite clearly undesirable effect can be easily experimentally observed. It can result in significant spread of the device characteristics and sensitivity to mechanical vibration.

The serpentine spring design effectively solves this problem. In fact, much higher residual stress levels, up to at least 100 MPa compressive, do not show any appreciable effect in simulations and no degradation attributable to stress was ever observed experimentally in these devices.

Design of serpentine and other periodic springs can be easily accomplished. Their stiffness, both torsional and translational, can be calculated based on the standard small displacement beam theory [4].

## DOUBLE HINGE TILTING MIRRORS

In the types of mirrors discussed above, electrostatic force is applied directly to the moving element. This imposes limitations on the actuation range and the drive voltage. It is sometimes more efficient to decouple the force-producing element from the moving element and connect them with a type of a transmission mechanism. This allows more flexibility in designing the force mechanism. For example, the voltage can be lowered by using a small-gap parallel-plate actuator and then converting the small displacement into the required large tilt. A "double-hinge" tilting mirror device shown in Fig 4 uses this actuation approach.

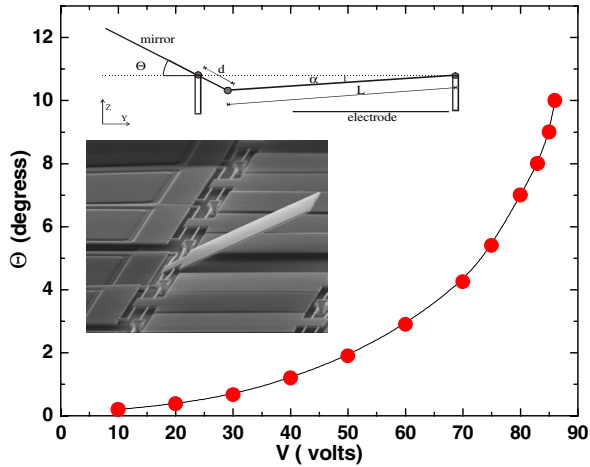

*Figure 4. Double Hinge tilting mirror. Mechanical tilt angle as a function of applied voltage. Bottom shows a micrograph of the device in the snap-down position. Top schematic illustrates the angle amplification principle.*

This approach relies on the effectiveness of the transmission mechanism, in this case the "double-hinge" connecting the actuator to the tilting reflector. Such mechanisms commonly require elements with significant difference in stiffness with respect to different deformations. Here torsional deformation in the hinges should be much easier to produce than linear displacement. More specifically $\tau / d < K$ where $\tau$ is the torsional stiffness, $K$ is the vertical stiffness and $d$ is the length of the lever arm. Note that $d$ should be kept small since it determines the degree of amplification.

This disparity in stiffness is easier to accomplish by a rocking-type mechanical contact point, which effectively has very low $\tau$ but a very large $K$. Mechanical contact effectively allows one to design extremely non-linear mechanical suspensions: there is a sharp discontinuity in the displacement as a function of force when contact occurs. In particular this allows one to design mechanical links (springs) soft for some deformations and very hard for others.

In MEMS actuator design however, the best performance and reliability is achieved with pure flexure elements, without rubbing mechanical joints or other surfaces in contact. Considering the pure flexure suspension, one is tempted to try and replicate the nonlinearities of the mechanical contact with designs like straight rods, similar to the one shown in Fig. 4. But such nonlinearities fundamentally come at a price of increased sensitivity to residual stress, since the nonlinear behavior manifests itself exactly when the stress (e.g., tension) is induced within the beam. As we have shown before, such designs may not be sufficiently robust to stress variations.

It is possible to design strain-relieved, linear elastic springs which possess the required stiffness ratio. Such springs are preferable from the standpoint of performance (repeatability, robustness) and reliability (wear). We have used the periodic serpentine spring analysis mentioned above to design the linear elastic type of the "double-hinge" joint. We have successfully fabricated such devices. Their performance is illustrated in Fig. 4. Large tilting angles are achieved at moderate voltages while operating in the highly repeatable, reliable and stable pure-flexure regime.

The implementation of such designs is critically dependent upon the availability of the process that can accurately create features much smaller than the characteristic feature size of the rest of the device, such as beams with high aspect ratio cross-sections.

## SELF-ASSEMBLY

While in the previous examples the effects of the residual stress had to be minimized, the next few examples show how the residual stress can be creatively utilized. Similarly to the way thermally induced stress powers thermal actuators, residual stress can produce useful work.

In surface micromachining and other processes many complex devices can not be produced in their final shape or form. Instead they have to be assembled – individual parts have to be moved to their final position and locked there. For example a shutter has to be locked in a vertical position, or a tilting mirror has to be positioned at a given height above the electrodes. The residual stress can be used to perform this assembly automatically. Such self-assembly is incorporated in a micromirror shown in Fig. 5. [5]

The self-assembly is accomplished during the release step of the processing sequence on all mirrors simultaneously without either human intervention or external power supply. The mechanical energy is stored during deposition in the special high-stress layer, which is put on top of the four assembly arms. Immediately after the assembly arms are released, the tensile stress in this layer causes them to bend up and push the mirror frame, lifting it in place above the Si substrate. The tapered cuts in the hinged sidewalls are engaged with the dovetail structures at the frame edge, and as the frame is raised, the sidewalls are rotated 90 degrees out of their initial position within the substrate plane. In their final, vertical position, the sidewalls provide a lithographically defined accurate stop, locking the frame precisely in place.

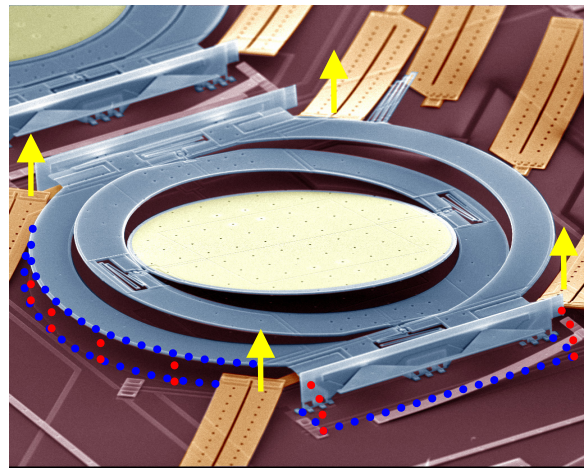

*Figure 5. Self-assembling beam-steering micromirror.*

To ensure stability of the frame and mirror position above the electrodes, the self-assembly holding force exceeds by a large margin any electrostatic and g-shock forces the device can be subjected to. A freestanding, unloaded bimorph self-assembly arm will assume the shape of an arc of radius R, given by the arm's layer thicknesses, elastic moduli and the residual stress of the materials. When the arm is loaded at the tip with a force F, its shape for small displacements is given by the superposition of the initial arc and the deformation due to the load:

$$z(x) = \frac{x^2}{2R} - \frac{Fx^2(3L-x)}{6EI},$$

where z is the position of the point on the arm at distance x from the origin, L is the length of the arm, E is the beam Young's modulus and I is the crossection moment of inertia.

If the assembly arm lifts the mirror to a height h, the holding force is given by:

$$h = z(L) \implies F = 3EI\frac{L^2/2R - h}{L^3},$$

which dependence is illustrated in Fig. 6.

For our self-assembling mirror the holding force produced by the four arms is in excess of 70uN, corresponding to the weight of the mirror structure during a ~2000g mechanical shock test. The maximum possible electrostatic force applied to the mirror does not exceed 10uN. Thus our design is quite stable.

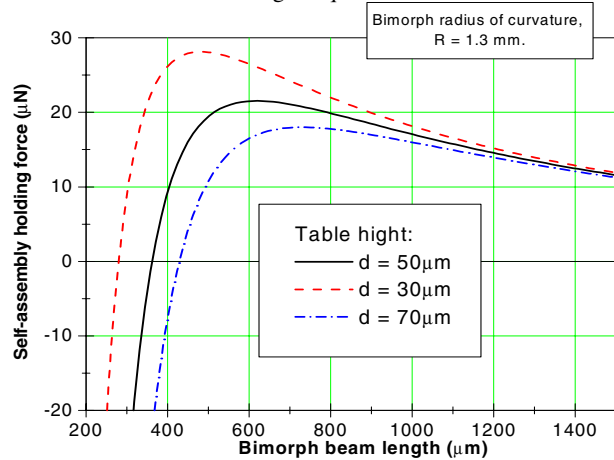

*Figure 6. Forces produced by a single self-assembly arm for different lifting heights and arm lengths.*

## PARTY-FAVOR TILTING MIRRORS

The device shown in Fig. 7 combines residual stress engineering with yet another electrostatic approach and functions as a piston-tilt micromirror. The residual stress is used to modify the shape of the actuator elements and position the optical device initially away from and at an angle to the substrate. The curled arm itself is used as a movable, elastic element of the actuator, uncurling itself gradually as voltage is increased on the fixed electrode directly beneath. As the arm uncurls, the legs with dimples at their edge come in contact with the grounded landing pads at the sides of the fixed electrode. The dimples on the legs protrude downward, lower than the bottom surface of the arm itself, defining the final minimum gap between the arm and the electrode underneath and preventing the arm from touching and shorting to the electrode.

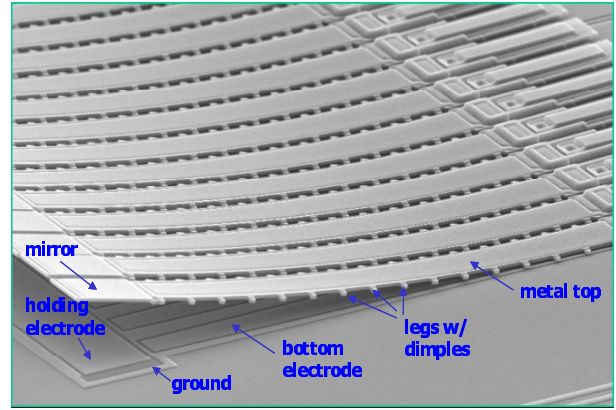

*Figure 7. "Party-favor" tilt-piston devices.*

This device is an example of a strongly coupled electrostatic-mechanical system, as the shape of a conductor rather than just its position, changes as a result of the applied electrostatic force. It also uses mechanical contact during operation. This is a static contact with normal load. When contact can not be avoided, the order of preference for design is first a fixed contact point with normal load, a fixed contact point with normal and lateral load and the last is a contact point sliding under normal and lateral load.

As shown in Fig. 8, these devices achieve tilt angles comparable to the ones attained by the Double Hinge devices at about the same actuation voltage. But here no fine lithography is required, and the actuation voltage is independent of the amplitude of motion (defined by the arm's length), allowing a significant voltage reduction.

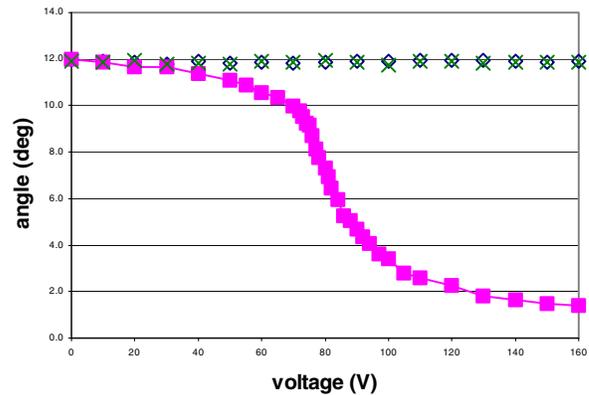

*Figure 8. "Party-favor" mirror tilt as a function of applied voltage. Curved line- device with voltage applied, straight lines- two nearest-neighbor devices, showing no crosstalk.*

Indeed, for a rough estimate we can neglect electrostatic forces where separation is large and only consider the region in the vicinity of the point of contact. The voltage required to start uncurling the arm is dependent only upon the arm stiffness and curvature, electrode width and the *smallest* gap between the electrode and the arm (defined by the dimple height). For example if the dimple height is reduced, the actuation voltage required for the same amplitude of motion is reduced in proportion with the gap. If the electrode is of a uniform width, the device shows a threshold behavior – when the voltage reaches the "uncurling" voltage, the device will uncurl completely. To make the uncurling motion gradual (~70V to ~100V in Fig. 8), the electrode is tapered, so that larger voltage is required to uncurl the last portion of the device than the first portion. Alternatively, the electrode can be made uniform and the curvature of the arm increasing toward the free end to achieve the same effect.

The slope of the curve from 0 to 70V is explained by the far-field part of the electrostatic force, and the slope above 100V by the fact that the electrode does not extend all the way under the free end on the arm.

## STRESS IN BIMORPH PLATES BI-STABLE ACTUATOR

The curvature arising in mirrors consisting of two or more material layers due to dissimilar stress is a widely discussed topic. Small isotropic stresses in uniform layers lead to spherical distortions in thin plates. Various ways of reducing such unwanted curvature have been demonstrated, from better stress control to stiffening the plates to stress compensation in multiple layer stacks.

It is interesting to take a look at the other limit, where the stress and the distortion are large. Unlike beams where the small displacement linear theory is applicable for displacements smaller than the beam length (with a few exceptions), for plates the problem becomes nonlinear at displacements comparable to the plate *thickness*. Indeed, such deformations generally cannot occur without considerable in-plane stress. For example a disc can be deformed to become spherical, but cannot form a complete sphere without being significantly stretched or compressed. An exception to the rule is the cylindrical deformation, which requires only bending and no in-plane stretching or compression.

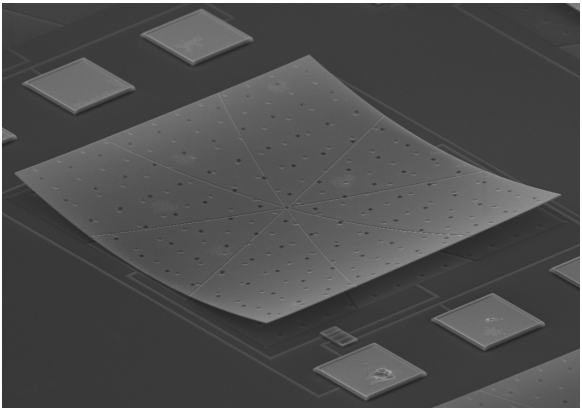

*Figure 9 Bi-stable actuator plate.*

If we consider a square bimorph plate with dissimilarly stressed materials, for small stresses the equilibrium shape of the plate is spherical. If we increase the stress, the radius of the sphere will linearly decrease up to a point when the deformation of the plate becomes comparable to plates thickness. If the stress is increased further, the deformation shape starts to deviate from a sphere until a point where a bifurcation occurs and the symmetry is spontaneously broken. As the stress is further increased the plate assumes cylindrical shape, as shown in Fig. 9.

Note however that there are two equivalent directions in which the axis of the cylinder may be pointing – along either of the edges of the plate. Thus this mechanical structure is bi-stable, with two equivalent equilibrium shapes. It is at the same time effectively strain relieved – its shape would not change if the substrate were stretched or compressed.

The plate in the figure can be switched between stable positions by pulling on it electrostatically with sets of electrodes underneath, similar to the "party-favor" device above. This device can be used for building pure-flexure latchable low voltage out of plane actuators.

## CONCLUSIONS

Although the examples above mostly are tilting micromirrors of various types, different optical elements and actuators can be and have been used to create a wide variety of optical MEMS devices. Scaling down of existing macromechanical optics provides a way toward cheaper, smaller and faster optical components, while maintaining excellent optical performance: low insertion loss, high contrast, low polarization and low wavelength dependence, etc. More importantly, dense integration of MOEMS devices on a single chip enables completely new optical subsystems, where large numbers of optical signals can be switched or manipulated simultaneously.

For such applications electrostatic actuation principles provide low power dissipation, low crosstalk, excellent performance and design flexibility and ease of fabrication. Although with conventional approaches actuation voltage tends to increase with increased mechanical range, unconventional types of actuators allow circumventing the problem without compromising speed or mechanical robustness.

Design of elastic elements and residual stress engineering are the other two important themes in actuator design for optical MEMS, detailed understanding and use of which leads to robust, reliable, high-performance devices.

Nonlinear phenomena, mechanical, electrostatic and coupled, are abundant in MEMS actuators. Such phenomena can be effectively avoided or harnessed for productive purposes using various sorts of analytical and numerical analysis techniques.

## REFERENCES

1. D. J. Bishop, C. R. Giles, and S. R. Das, "The Rise of Optical Switching", *Scientific American*, Jan 2001, pp. 88-94.
2. A. Neukermans, 'MEMS devices for all optical networks' *Proc. SPIE* Vol. 4561 *MOEMS and Miniaturized Systems II*, pp.1-10, 2001.
3. V. Aksyuk, B. Barber, C.R. Giles, R. Ruel, L. Stulz, and D. Bishop, "Low Insertion Loss Packaged and Fiber-connectorized Si Surface-Micromachined Reflective Optical Switch," *Solid-State Sensor and Actuator Workshop*, pp. 79-82, Hilton Head Island, South Carolina, June 8-11, 1998.
4. L. D. Landau and E. M. Lifshitz, *Theory of Elasticity, volume 7 of the course of Theoretical Physics,* translated by J. Sykes and W. Reid (Addison-Wesley Publishing Company, Inc., Reading, Massachusetts, 1959).
5. V. A. Aksyuk, F. Pardo, D. J. Bishop, "Stress-induced curvature engineering in surface-micromachined devices", Proc. SPIE, vol. 3680, p. 984, March 1999.